\documentclass[
    ,final            
  ]
  {aipproc}

\layoutstyle{6x9}

\newcommand{\lf}{\left}
\newcommand{\rg}{\right}

\newcommand{\AmS}{{\protect\the\textfont2
 A\kern-.1667em\lower.5ex\hbox{M}\kern-.125emS}}
\usepackage{amssymb}
\usepackage{epsfig}

\newcommand{\bm}[1]{\mbox{\boldmath $#1$}}

\newcommand{\FF}{{\mathcal F}}

\newcommand{\bea}{\begin{eqnarray}}
\newcommand{\eea}{\end{eqnarray}}

\newcommand{\nn}{\nonumber \\}
\newcommand{\nnn}{\nonumber}
\newcommand{\be}{\begin{eqnarray}}
\newcommand{\ba}{\begin{array}}
\newcommand{\ea}{\end{array}}
\newcommand{\ee}{\end{eqnarray}}

\newcommand{\Tr}{{\rm Tr}}

\newcommand{\h}{\hat{\bm{h}}}

\begin{document}

\title{Transversity Properties of Quarks and Hadrons in SIDIS and Drell-Yan}


\author{Leonard P. Gamberg}{
  address={Division of Science, Penn State Berks, Reading, PA 19610, USA}
}

\author{Gary R. Goldstein}{
  address={Department of Physics and Astronomy, Tufts University, Medford,
MA 02155, USA}
}

\begin{abstract}
We consider the leading
twist $T$-odd  contributions as  the dominant source of the
azimuthal and transverse single spin asymmetries in SIDIS and 
dilepton production in Drell-Yan Scattering.
 These asymmetries contain
information on the distribution of quark transverse spin
in  (un)polarized protons.
In the spectator framework we estimate 
these asymmetries at HERMES kinematics and 
at $50\ {\rm GeV}$ 
for the proposed experiments 
at GSI, where an anti-proton beam is ideal for studying the
transversity properties of quarks due to the dominance of {\em valence} quark
effects.
\end{abstract}

\maketitle


One of the persistent challenges confronting the QCD
parton model is to provide a theoretical basis for
the experimentally 
significant azimuthal and transverse
spin asymmetries that emerge in  inclusive and semi-inclusive
processes.  Generally speaking, the spin dependent amplitudes for the scattering will
contribute to 
non-zero transverse single spin asymmetries (SSA) if there are imaginary parts of
bilinear products of those amplitudes that have overall helicity change. 
In perturbative QCD (PQCD), applicable to the hard scattering region, to obtain
an imaginary contribution to quark and/or gluon scattering processes demands introducing higher order
corrections to tree level processes. One approach incorporates the requisite phases
through interference of tree level and one-loop contributions in PQCD in an
attempt to explain up-down polarization 
asymmetry in $\Lambda$ production~\cite{gandd}. On
general grounds  the helicity conservation
property of massless QCD predicts that such contributions are  small,
going like $\alpha_s m/Q$, where $\alpha_s$ is the strong coupling, $m$ represents a non-zero quark mass
and $Q$ represents the hard QCD scale~\cite{gandd,kane}. Such contributions have failed
to account for the large SSA observed in $\Lambda$ production~\cite{heller}. 

However, considering the soft contributions to hadronic
processes opens up the possibility that there are
non-trivial transversity parton distributions that can
contribute to transverse spin asymmetries~\cite{RS}.  
For transverse  SSA in SIDIS,
transverse momentum must be acquired to lead to
appropriate helicity changes at leading twist. 
In describing transverse asymmetries
this is particularly relevant when
the transverse momentum can arise from  
intrinsic quark momenta.  Here the
effects are associated with non-perturbative transverse
momentum distribution functions~\cite{soper} (TMD),
where transverse SSAs indicate  so called $T$-odd correlations between
transverse spin and longitudinal and intrinsic quark transverse momentum.
The  $T$-odd distributions~\cite{sivers,bm}
are  of importance as they possess both transversity properties and the
necessary phases to account for SSA and azimuthal 
asymmetries~\cite{colnpb,bhs}.  
Formally, these phases can be generated from the
gauge invariant definitions of the $T$-odd
quark distribution functions~\cite{colplb,jiyuan,gg}.
In contrast to the  transverse SSAs generated from the interference of
tree-level and one loop correction in PQCD, 
such effects go like $\alpha_s {<}k_\perp {>}/M$,
where now $M$ plays the role of the chiral symmetry breaking scale and
$k_\perp$ is characteristic of quark intrinsic motion.


Here we consider the leading
twist $T$-odd  contributions as  the dominant source of the 
$\cos 2\phi$ azimuthal asymmetry and $\sin (\phi\pm\phi_s)$ transverse
SSAs in  SIDIS~\cite{ggoprh} and
azimuthal asymmetry $\nu$ in 
dilepton production in Drell-Yan Scattering~\cite{ggdy}. 
Among other interesting
properties, these asymmetries 
contain information on the distribution of quark transverse spin
in an unpolarized proton, $h_1^{\perp}(x,k_\perp)$~\cite{bm}.
In a parton-spectator framework we estimate 
these asymmetries  at HERMES kinematics~\cite{hermes} and for Drell-Yan
scattering at $50\ {\rm GeV}$  center of mass energy. 
The latter  azimuthal asymmetry  
is interesting in light of proposed 
experiments at GSI, where an anti-proton beam will ideal for studying the
transversity properties of quarks due to the dominance of {\em valence} quark
effects~\cite{pax}.

The leading twist contributions to the factorized cross-section  
for a transversely polarized nucleon target in lepton-proton scattering are
{\small
\be
&&\frac{d^6\!\sigma_{UT}^{\ell N^{\uparrow}\rightarrow \ell\pi X}}
{dx_H dy dz_h d\phi_S d^2\! \bm{P}_{h\perp}} = 
\frac{2 \alpha^2 }{Q^2 y}\,\Bigg\{
|\bm{S}_T|(1-y)\sin\left(\phi_h +\phi_S\rg)\sum_q e_q^2\,
 {\FF}\lf[\frac{\bm{p}_\perp\cdot\h}{M_h}\, 
h_{1}^{q} H_1^{\perp q} \right] 
\nn && \hskip 1.5cm
+ \, |\bm{S}_T|\frac{\left(1+(1-y)^2\right)}{2} \sin\lf(\phi_h -\phi_S\rg)\sum_q e_q^2\, 
{\FF}\lf[\frac{\bm{k}_\perp\cdot\h}{M}\, f_{1T}^{\perp q} D_1^q \rg]\Bigg\} ,
\label{e:dUT}
\ee}
where $\FF$ is the convolution integral~\cite{bm}.  
The  twist two $T$-even and odd distribution and fragmentation functions  
appearing in Eq.~(\ref{e:dUT}) are
projected from the correlation functions for the
transverse momentum dependent distribution and fragmentation correlators,  
$\Phi(x,P)$ and $\Delta(p,P_h)$ respectively,
{\small
\bea
\Phi(x,\bm p_\perp)\hspace{-.2cm}&=&\hspace{-.2cm}{\frac{1}{2}}
\Bigl\{
f_1(x,\bm p_\perp){\not{n_+}} +
h_1^{\perp}(x,\bm p_\perp)\frac{{\sigma}_{\mu\nu}
p_{\perp}^{\mu}n^\nu_{+}}{M} 
+f_{1T}^\perp(x,\bm p_\perp)\frac{\epsilon_{\mu\nu\rho\sigma}\gamma^\mu n^\nu_+ p_\perp^\rho S_T^\sigma}{M}
\cdots \Bigr\}\ 
\nn
\Delta(z,\bm k_\perp) \hspace{-.2cm}&=&\hspace{-.2cm}
\frac{1}{4}\Bigl\{ D_1(z,z\bm k_\perp){\not{n_-}} +
H_1^{\perp}(z,z\bm k_\perp)
\frac{{\sigma}_{\mu\nu}k_{\perp}^\mu n^\nu_{-}}{M_h}
+\cdots \Bigr\},
\nnn
\eea}
where for example 
$\int dp^-\Tr\left(\sigma^{\perp +}\gamma_5\Phi\right)
=
\frac{2\varepsilon_{\scriptscriptstyle
+-\perp j}\, p_{\scriptscriptstyle \perp j}}{M}h_1^\perp(x,p_\perp)
\dots\quad .$
We use the parton inspired quark-diquark spectator
framework to model the quark-hadron interactions
that enter the $T$-odd and even TMDs and fragmentation functions
contributing to  $\Phi(x,P)$ and $\Delta(z,P_h)$~\cite{ggoprh}.
Noting that parton intrinsic transverse 
momentum yields a natural  regularization for the moments of 
these distributions, we incorporated a Gaussian from factor  
into our model.   The resulting  scalar diquark contribution is
{\small $h_1^\perp(x,p_\perp)=
{\cal N}\alpha_s M
\frac{(1-x)(m+xM)}{p_\perp^2\Lambda(p^2_\perp)}{\cal R}(p_\perp^2;x)$}
where 
{\small${\cal R}(p_\perp^2;x)=
\exp^{-2b(p^2_\perp-
\Lambda(0))}\left(\Gamma(0,2b \Lambda(0))-
\Gamma(0,2b \Lambda(p^2_\perp))\right)$}
is the regularization function.  
$\Lambda(k_\perp^2)$ is the spectral function and
${\cal N}$ is a normalization factor determined 
with respect to   the unpolarized $u$-quark distribution, 
obtained from the  zeroth moment of $f_1^{(u)}(x,p_\perp)$
normalized with respect to valence distributions.
Our regulated expression of the Collins function is given by
{\small$H_1^\perp(z,k_\perp)=
{\cal N}^\prime\alpha_s 
\frac{1}{4z}\frac{(1-z)}{z}
\frac{\mu}{\Lambda^\prime(k^2_\perp)}
\frac{M_\pi}{k_\perp^2}{\cal R} (z,\bm{k}_\perp^2)$}
where $\mu$ is the quark spectator mass
and $\cal N^\prime$ is determined from the normalization on the
unpolarized fragmentation function $D_1(z)$.
The Collins and Sivers  weighted asymmetries 
are projected from the differential cross sections, Eq.~(\ref{e:dUT})
{\small\bea
\langle \frac{P_{h\perp}}{M_\pi}
\sin(\phi+\phi_s) \rangle_{\scriptscriptstyle UT}
\hspace{-.3cm}&=&\hspace{-.3cm}\frac{\int \hspace{-0.1cm} d\phi_s  
{ d^2P_{h\perp}}\frac{P_{h\perp}}{M_\pi}
\sin(\phi\hspace{-0.1cm}+\hspace{-0.1cm}\phi_s)
\left(d\sigma^{\uparrow}\hspace{-0.1cm}-\hspace{-0.1cm}
d\sigma^{\downarrow}\right)}
{\int d\phi_s\int d^2P_{h\perp}\left(d\sigma^{\uparrow}
+d\sigma^{\downarrow}\right)}
\hspace{-.1cm}=\hspace{-.1cm}\frac{\big|S_T\big|2(1-y) \sum_q e^2_q h_1(x) z H^{\perp(1)}_1(z)}
{{(1+{(1-y)}^2)}  \sum_q e^2_q f_1(x) D_1(z)}
\nnn
\eea}
and 
$\langle \frac{\vert P_{h\perp} \vert}{M} 
\sin(\phi-\phi_S) 
\rangle_{\scriptscriptstyle UT}=
 \vert {\bm S_T} \vert\frac{(1+{(1-y)}^2) 
\sum_q e^2_q f^{\perp(1)}_{1T}(x) z D^q_1(z)}
{{(1+{(1-y)}^2)}  \sum_q e^2_q f_1(x) D_1(z)}$.
\begin{figure}[t]
  \includegraphics[width=0.25\textwidth,height=.14\textheight]{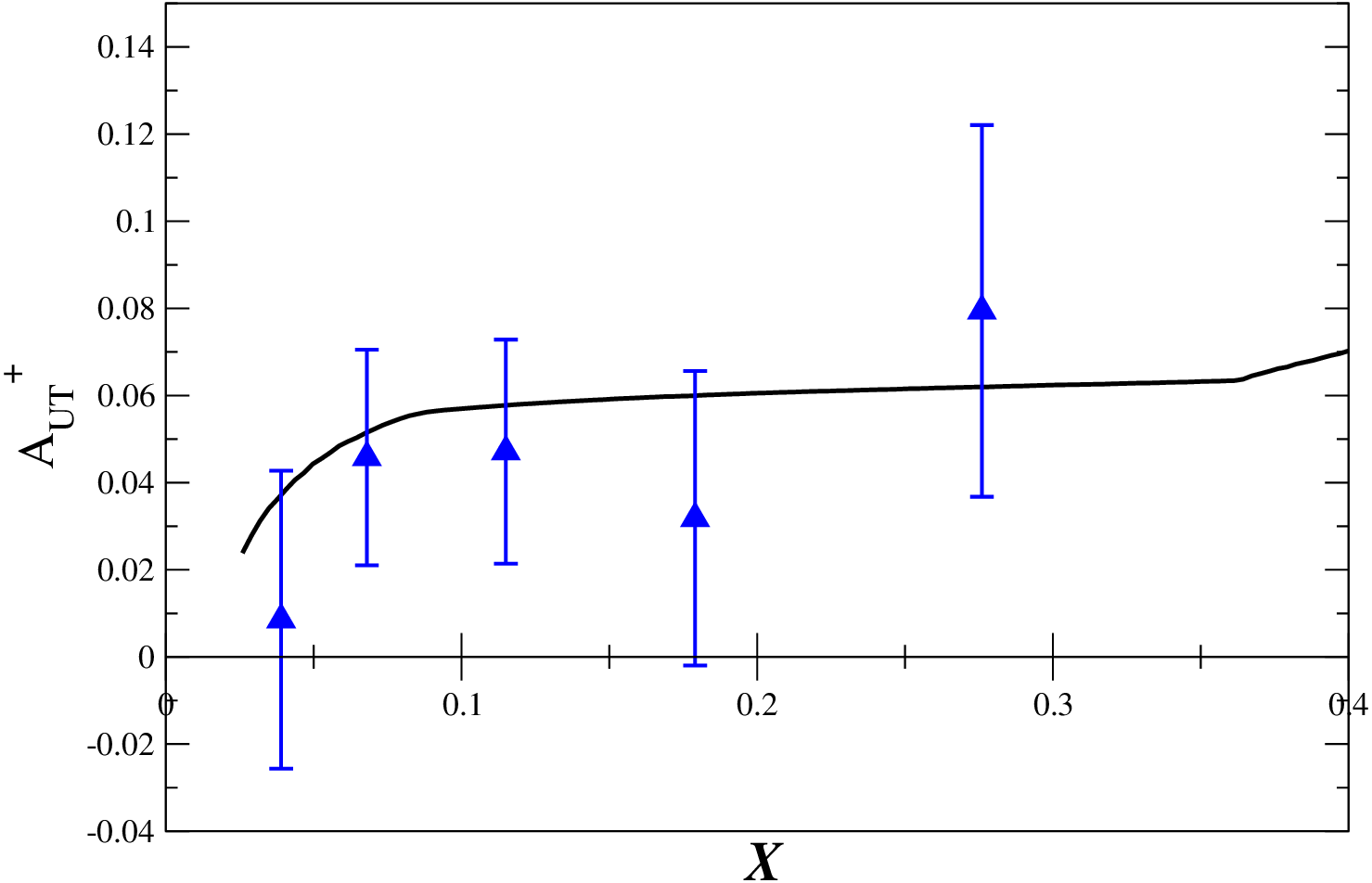}
  \includegraphics[width=0.25\textwidth,height=.14\textheight]{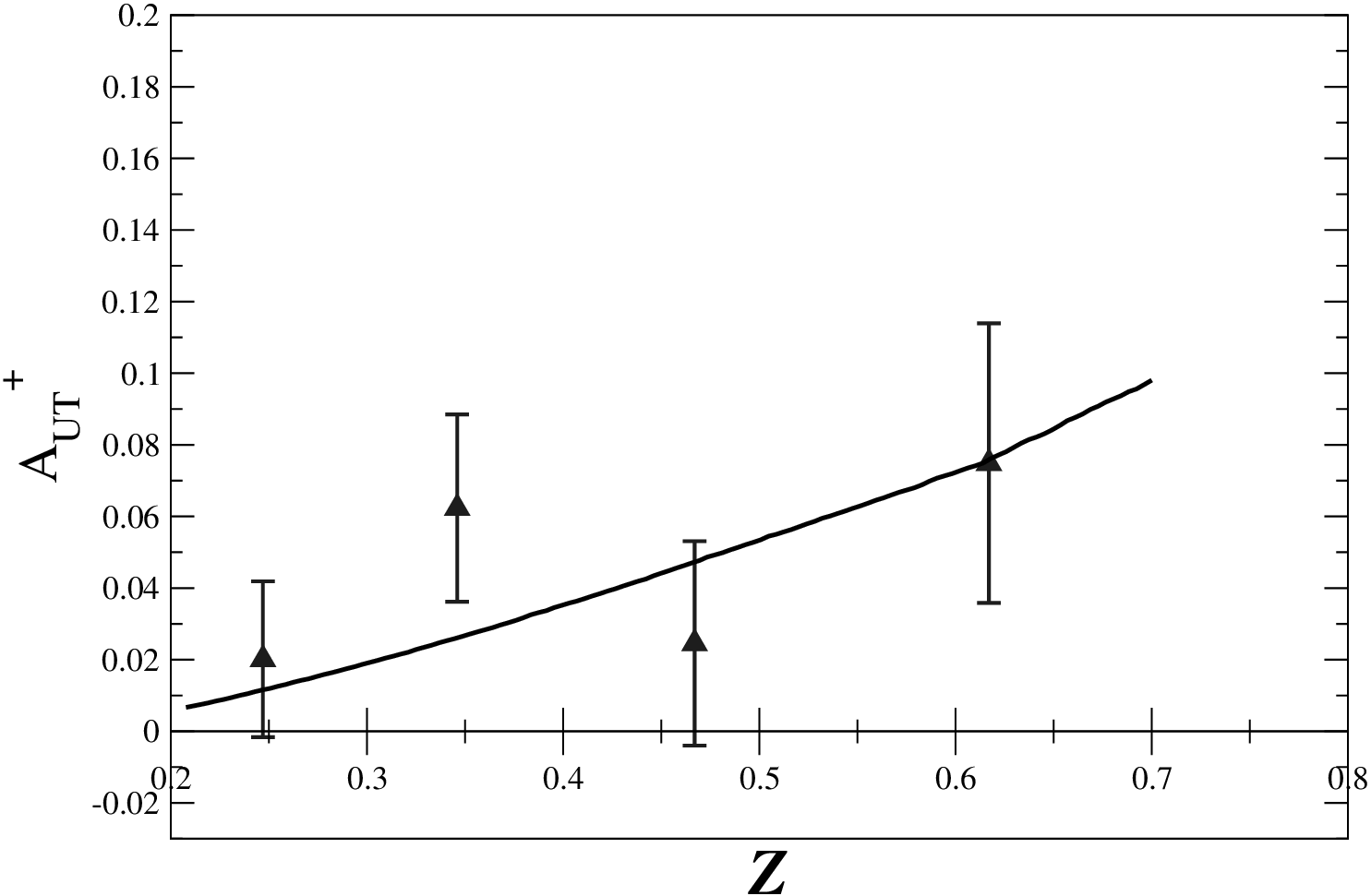}
  \includegraphics[width=0.25\textwidth,height=.14\textheight]{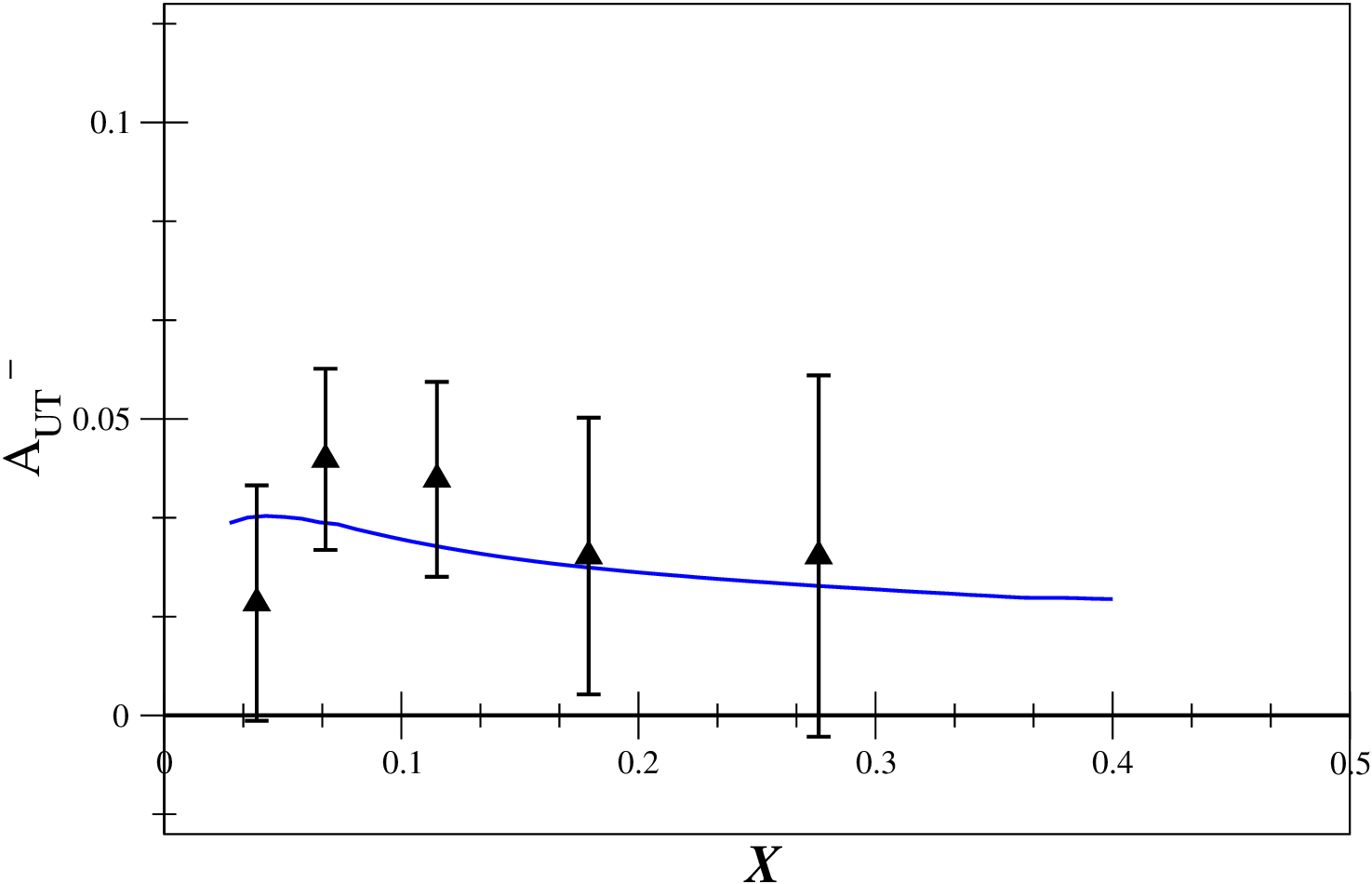}
  \includegraphics[width=0.25\textwidth,height=.14\textheight]{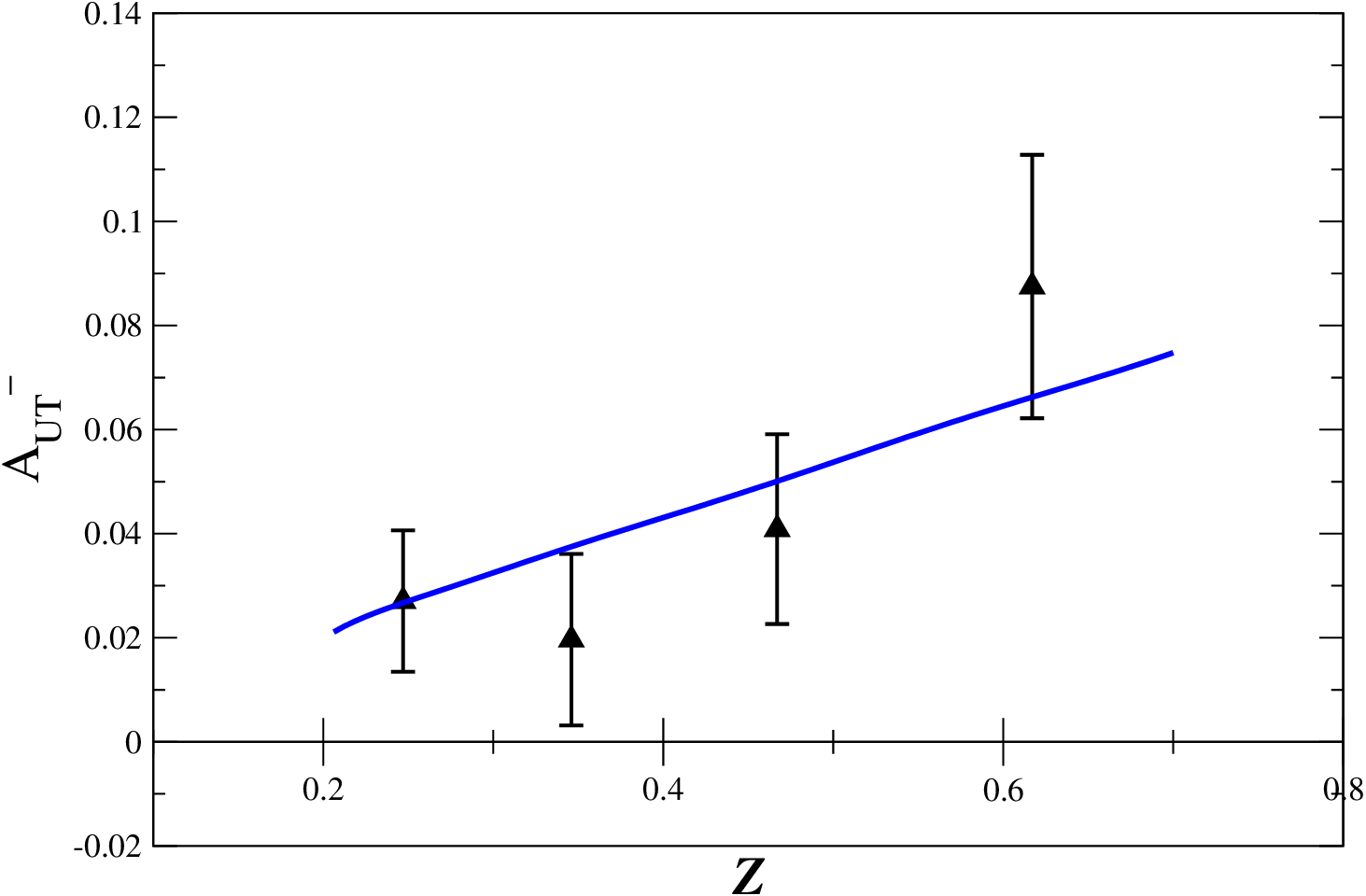}
  \caption{{ Left two Panel: The 
\protect{$\langle \sin(\phi+\phi_s) \rangle_{\scriptscriptstyle UT}$ } 
asymmetry for $\pi^+$ production as a function of $x$  and $z$ compared
to the HERMES data~\cite{hermes}.  Right two Panels: 
The $\langle\sin(\phi-\phi_S)\rangle_{\scriptscriptstyle UT}$ 
as a function of $x$ and $z$.}}
\label{fig1}
\end{figure}
We have re-analyzed these asymmetries including both the scalar and vector
diquark contributions to the TMDs for the central
values of our parameter set,
and compared the transverse SSAs to the 
the HERMES data~\cite{hermes} for $\pi^+$ production in Fig.~\ref{fig1}.
The unweighted asymmetries are approximated as
{\small
$A_{UT}^{\sin(\phi+\phi_s)}\approx\frac{M_\pi}{\langle P_{h\perp}\rangle}
\langle\frac{P_{h\perp}}{M_\pi}\sin\left(\phi+\phi_s\right)\rangle
\quad{\rm and }\quad
A_{UT}^{\sin(\phi-\phi_s)}\approx\frac{M}{\langle P_{h\perp}\rangle}
\langle\frac{P_{h\perp}}{M}\sin\left(\phi\pm\phi_s\right)\rangle$}.  
These results agree to within the errors displayed.
\begin{figure}[t]
  \includegraphics[width=0.23\textwidth,height=.14\textheight]{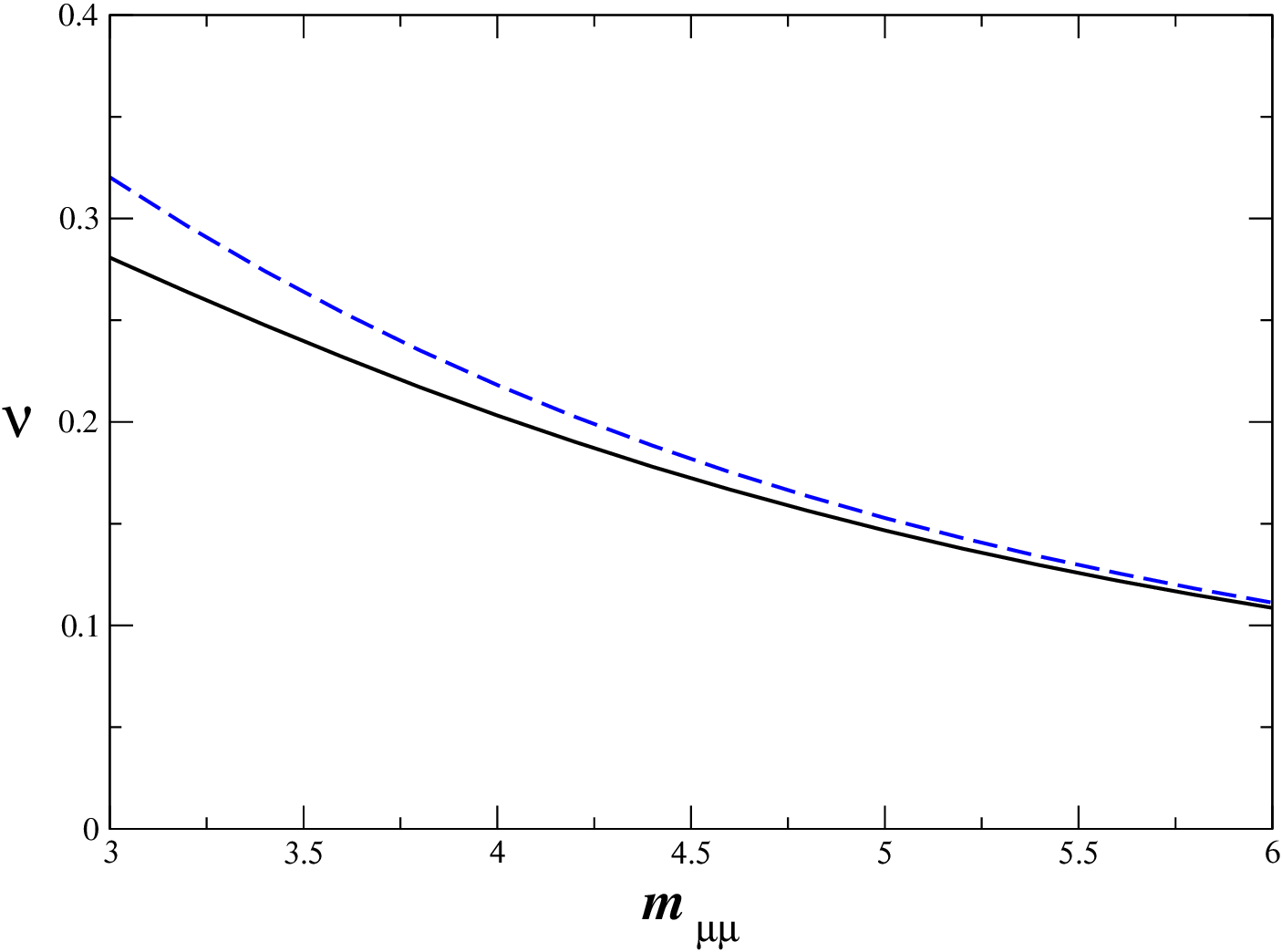}\hskip .35cm
  \includegraphics[width=0.23\textwidth,height=.14\textheight]{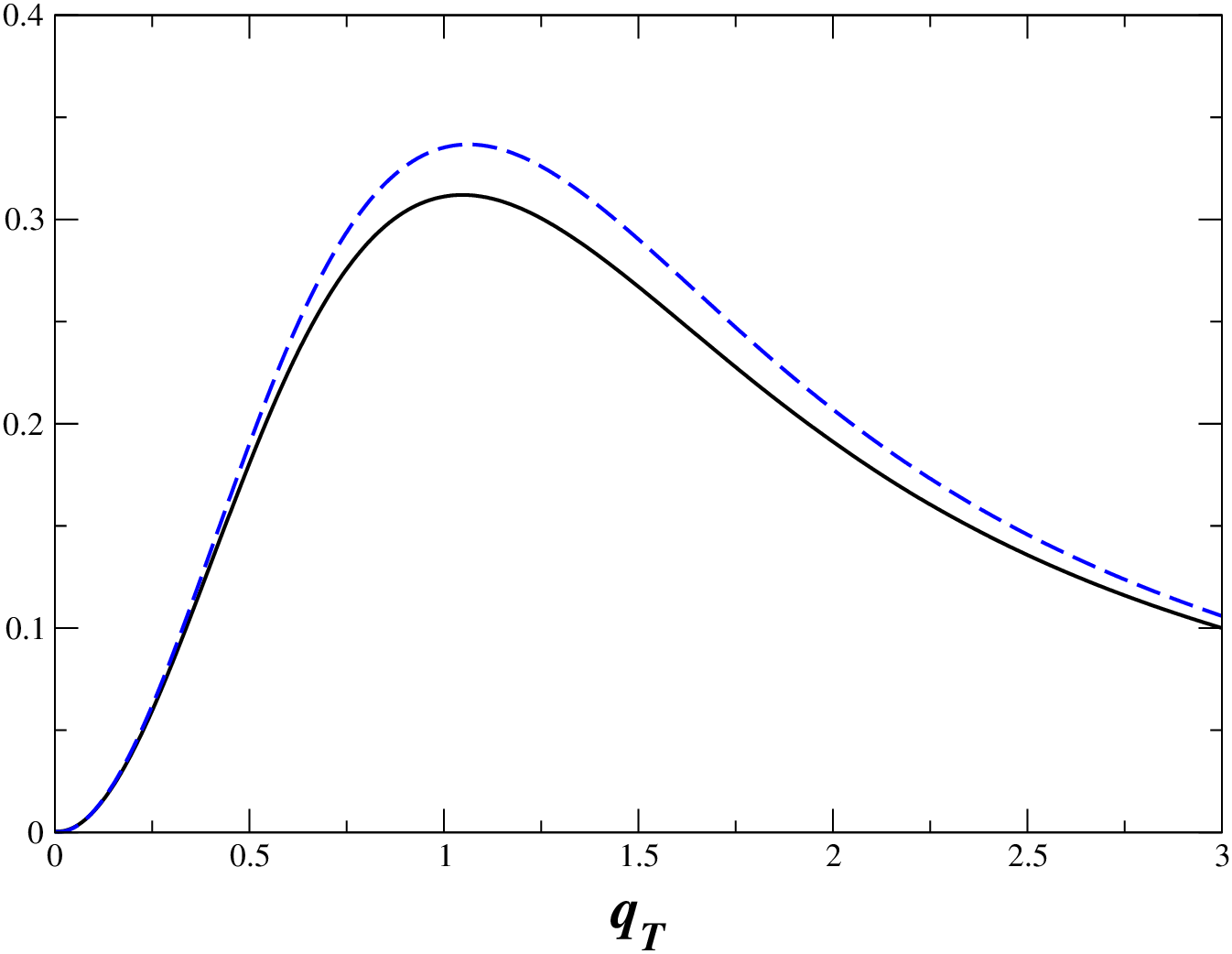}\hskip .35cm
  \includegraphics[width=0.23\textwidth,height=.14\textheight]{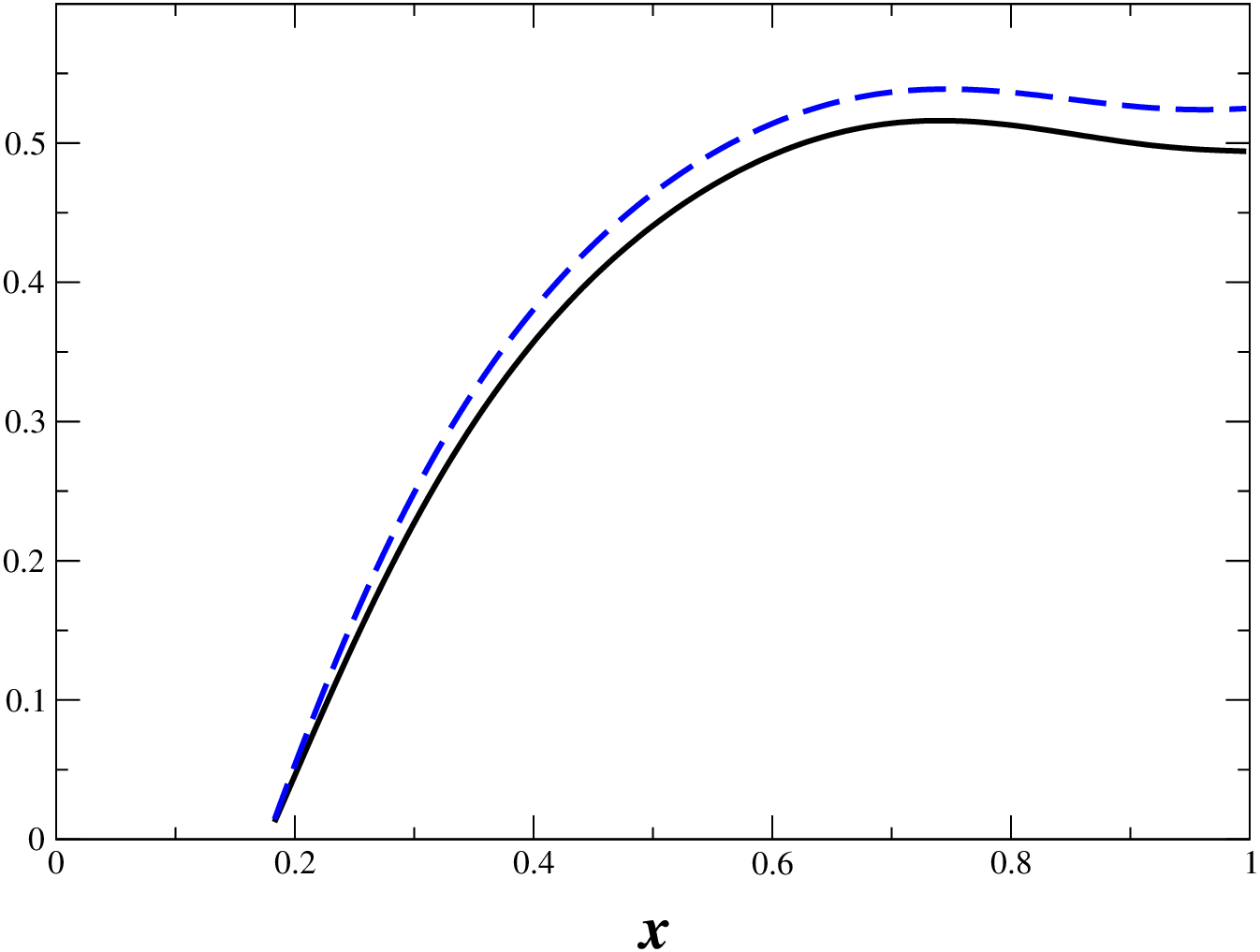}\hskip .5cm
  \includegraphics[width=0.23\textwidth,height=.14\textheight]{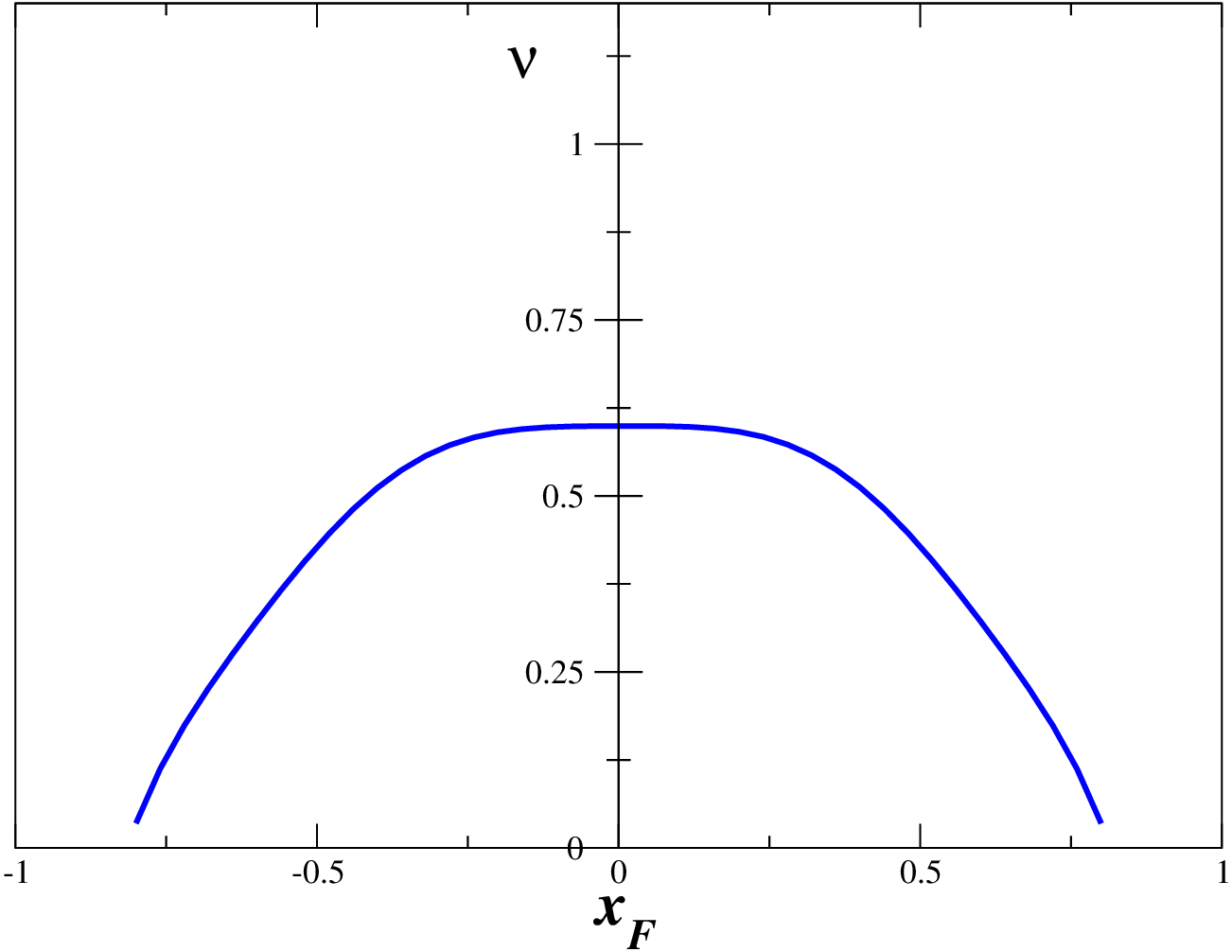}
\caption{{ Left two Panels:
$\nu$ plotted as a function of $q_T$ and $q=m_{\mu\mu}$ for
$s=50\ {\rm GeV}^2$, $x$ in the range $0.2-1.0$.  Right two panels:
$\nu$ plotted as a function of $x$ and $x_F$ for 
$s=50\ {\rm GeV}^2$ $q_T$ ranging from 3 to 6 GeV/c and $q$ from
0 to 3 GeV/c.}} 
\label{fig2}
\end{figure}

An  unpolarized double 
$T$-odd azimuthal asymmetry enters the 
Drell-Yan process~\cite{boerich}.  
For the Drell-Yan process the angular dependence~\cite{cs}
can be expressed as
{\small
\bea
\frac{d N}{d\Omega}\hspace*{-.25cm}&\equiv &\hspace*{-.25cm}
=
\frac{3}{4\pi}\frac{1}
{\lambda+3}\Big( 1+\lambda\cos^2\theta \hspace{-.05cm}+\hspace{-.05cm}\mu \sin^2\theta 
\cos\phi
+\frac{\nu}{2} \sin^2\theta \cos 2\phi\Big),
\label{cross}
\eea}
where {\small$\frac{d N}{d\Omega}\equiv
\left(\frac{d\sigma}{dQ^2 dy d\bm{q}_T^2}\right)^{-1}
\frac{d\sigma}{dQ^2 dy d\bm{q}^2_T d\Omega}$}.  The solid angle $\Omega$ 
refers to the lepton pair orientation in 
the pair rest frame relative to the boost direction,
and   $\lambda, \mu, \nu$ are functions that depend 
on  $x, m_{\mu\mu}^2, \bm{q}_T$,
the fraction of quark momentum in the hadron,
the invariant mass of the produced lepton pair, and the transverse
momentum of the dimuon pair.  
All of the asymmetry functions, $\mu, \lambda$ and $\nu$,
  have parton model contributions which at next to leading order predict
$1-\lambda-2\nu=0$, the so called  Lam-Tung relation~\cite{lam}.
Experimental measurements of $\pi p \rightarrow \mu^+  \mu^- X$ discovered
unexpectedly large values of these asymmetries~\cite{e615} compared to 
parton-model expectations resulting in a serious violation of this relation.
It has been suggested \cite{boerich} 
that there is a dominant leading twist contribution to $\nu$ coming
from the $T$-odd transversity distributions $h_1^{\perp}(x,k_\perp)$ for
both hadrons which dominates in the kinematic range, $\bm{q_T}\ll Q$.
The $\cos 2\phi$ azimuthal asymmetry in  
unpolarized $p\, \bar{p}\rightarrow \mu^+\, \mu^-\, X$
involves the convolution of the leading twist $T$-odd  function, $h_1^\perp$
{\small
$\nu_2 =\frac{\sum_a e^2_a {\FF}
\left[w_2\,
h_1^\perp(x, k_\perp)\bar{h}_1^\perp(\bar{x}, p_\perp)/
(M_1 M_2)\right]
}{\sum_a e_a^2 {\FF}
\left[f_1(x, k_\perp) \bar{f}_1(\bar{x}, p_\perp)\right]}$}
where $w_2=(2 \hat{\bm{h}}\cdot \bm{k}_{\perp }\cdot\hat{\bm{h}}\cdot
\bm{p}_{\perp }
- \bm{p}_\perp \cdot \bm{k}_\perp)$
is the weight in the convolution
integral, ${\FF}$.  In addition it is known that
there is a non-leading
$T$-even contribution to the $\cos2\phi$
asymmetry\cite{cs}
{\small
$\nu_4=\frac{\frac{1}{Q^2}\sum_a e^2_a{\FF}
\left[w_4\,
f_1(x, k_\perp) \bar{f}_1(\bar{x}, p_\perp)
\right]}
{\sum_a e_a^2 {\FF}\left( f_1(x, k_\perp)
\bar{f}_1(\bar{x}, p_\perp)\right)},
$}
where
$w_4=2(\hat{\bm{h}}\cdot(\bm{k}_\perp -\bm{p}_\perp ))^2
-(\bm{k}_\perp-\bm{p}_\perp)^2$
and $\hat{\bm{h}}=\bm{q}_T/Q_T$. 
Fig.~\ref{fig2}
shows that the $\cos 2\phi$ azimuthal asymmetry $\nu$ 
is not small at center of mass energies of $50\ {\rm GeV}^2$.
However the $T$-odd portion dominates with an
additional $3-5\%$ from the
sub-leading  $T$-even piece.
Thus, aside from the
competing $T$-even effect,  the experimental
observation of a strong $x$-dependence
would indicate the presence of $T$-odd structures in {\it unpolarized} 
Drell-Yan scattering, 
implying that novel transversity properties of the  nucleon can be 
accessed  {\em without invoking beam or 
target polarization}.    
\vskip .15cm
  G.R.G. is supported by U.S. DOE (DE-FG02-92ER40702).



\end{document}